\documentclass[aps,prl,twocolumn,showpacs,preprintnumbers,amsmath,amssymb,superscriptaddress]{revtex4}
\usepackage{mathptm}  
\usepackage{dcolumn}                    
\usepackage{bm}                        
\usepackage{graphicx}
\usepackage{times}
\usepackage{epstopdf}
\begin{document}

%% Math
\newcommand{\Imag}{{\Im\mathrm{m}}}   % Imaginary part 
\newcommand{\Real}{{\mathrm{Re}}}   % Real part
\newcommand{\im}{\mathrm{i}}        % Imaginary unit non-italic
\newcommand{\talpha}{\tilde{\alpha}}
\newcommand{\ve}[1]{{\mathbf{#1}}}

\newcommand{\x}{\lambda}  % holder for plus/minus 1 (\pm 1)
\newcommand{\y}{\rho}     % holder for plus/minus 1 (\pm 1)
\newcommand{\T}{\mathrm{T}}   % Time ordering operator
\newcommand{\Pv}{\mathcal{P}} % Principal value operator
\newcommand{\vk}{\ve{k}} % Vector k
\newcommand{\vp}{\ve{p}} % Vector p

\newcommand{\N}{\underline{\mathcal{N}}} % Vector k
\newcommand{\Nt}{\underline{\tilde{\mathcal{N}}}} % Vector p
\newcommand{\g}{\underline{\gamma}} % Vector k
\newcommand{\gt}{\underline{\tilde{\gamma}}} % Vector p

\newcommand{\vecr}{\ve{r}} % Vector p
\newcommand{\vq}{\ve{q}} % Vector q
\newcommand{\ca}[2][]{c_{#2}^{\vphantom{\dagger}#1}} % op. c (annihilate) 
\newcommand{\cc}[2][]{c_{#2}^{{\dagger}#1}}          % op. c dagger (create) 
\newcommand{\da}[2][]{d_{#2}^{\vphantom{\dagger}#1}} % op. d (annihilate) 
\newcommand{\dc}[2][]{d_{#2}^{{\dagger}#1}}          % op. d dagger (create) 
\newcommand{\ga}[2][]{\gamma_{#2}^{\vphantom{\dagger}#1}} % op. gamma
\newcommand{\gc}[2][]{\gamma_{#2}^{{\dagger}#1}}          % op. gamma dagger 
\newcommand{\ea}[2][]{\eta_{#2}^{\vphantom{\dagger}#1}} % op. eta
\newcommand{\ec}[2][]{\eta_{#2}^{{\dagger}#1}}          % op. eta dagger 
\newcommand{\su}{\uparrow}    % Make the code more readable...
\newcommand{\sd}{\downarrow}  % Make the code more readable...
\newcommand{\Tkp}[1]{T_{\vk\vp#1}}  % Tunneling matrix element
\newcommand{\muone}{\mu^{(1)}}      % Chem.pot. side one
\newcommand{\mutwo}{\mu^{(2)}}      % Chem.pot. side two
\newcommand{\epsk}{\varepsilon_\vk}
\newcommand{\epsp}{\varepsilon_\vp}
\newcommand{\e}[1]{\mathrm{e}^{#1}}
\newcommand{\dif}{\mathrm{d}} %Rett d i differensial
\newcommand{\diff}[2]{\frac{\dif #1}{\dif #2}}%Derivert
\newcommand{\pdiff}[2]{\frac{\partial #1}{\partial #2}}%Derivert
\newcommand{\mean}[1]{\langle#1\rangle}
\newcommand{\abs}[1]{|#1|}
\newcommand{\abss}[1]{|#1|^2}
\newcommand{\Sk}[1][\vk]{\ve{S}_{#1}}
\newcommand{\pauli}[1][\alpha\beta]{\boldsymbol{\sigma}_{#1}^{\vphantom{\dagger}}}

%% Text
\newcommand{\eq}{Eq.}%No extra space when used with reftex (->auto ~)
\newcommand{\eqs}{Eqs.}%No extra space when used with reftex (->auto ~)
\newcommand{\cf}{\textit{cf. }}%adv : that is to say; in other words
\newcommand{\ie}{\textit{i.e. }}%adv : that is to say; in other words
\newcommand{\eg}{\textit{e.g. }}%[syn: f.eks., for example, for instance]
\newcommand{\etal}{\emph{et al.}}
\def\i{\mathrm{i}}

\title{Superconducting Proximity Effect in Silicene: Spin-Valley Polarized Andreev Reflection, Non-Local Transport, and Supercurrent }

\author{Jacob Linder}
\affiliation{Department of Physics, Norwegian University of
Science and Technology, N-7491 Trondheim, Norway}

\author{Takehito Yokoyama}
\affiliation{Department of Physics, Tokyo Institute of Technology, Tokyo 152-8551, Japan}

\date{\today}

\begin{abstract}
We theoretically study the superconducting proximity effect in silicene, which features massive Dirac fermions with a tunable mass (band gap), and compute the conductance across a normal$\mid$superconductor (N$\mid$S) silicene junction, the non-local conductance of an N$\mid$S$\mid$N junction, and the supercurrent flowing in an S$\mid$N$\mid$S junction. It is demonstrated that the transport processes consisting of local and non-local Andreev reflection may be efficiently controlled via an external electric field owing to the buckled structure of silicene. In particular, we demonstrate that it is possible to obtain a fully spin-valley polarized crossed Andreev reflection process without any contamination of elastic cotunneling or local Andreev reflection, in stark contrast to ordinary metals. It is also shown that the supercurrent flowing in the S$\mid$N$\mid$S junction can be fully spin-valley polarized and that it is controllable by an external electric field.

\end{abstract}
\maketitle

With the advent of graphene \cite{review_graphene} and topological insulators \cite{review_TI}, the study of Dirac fermions in condensed matter systems \cite{nodal} has become one of the most active research fields in physics over the last decade. Condensed matter systems
with such a 'relativistic' electronic band-structure are
intriguing examples of low-energy emergent symmetries (in this case, Lorentz-invariance). This has led to a tremendous amount of interest in terms of possible application value as well as from a fundamental physics viewpoint. 

\begin{figure}[t!]
\centering
\resizebox{0.48\textwidth}{!}{
\includegraphics{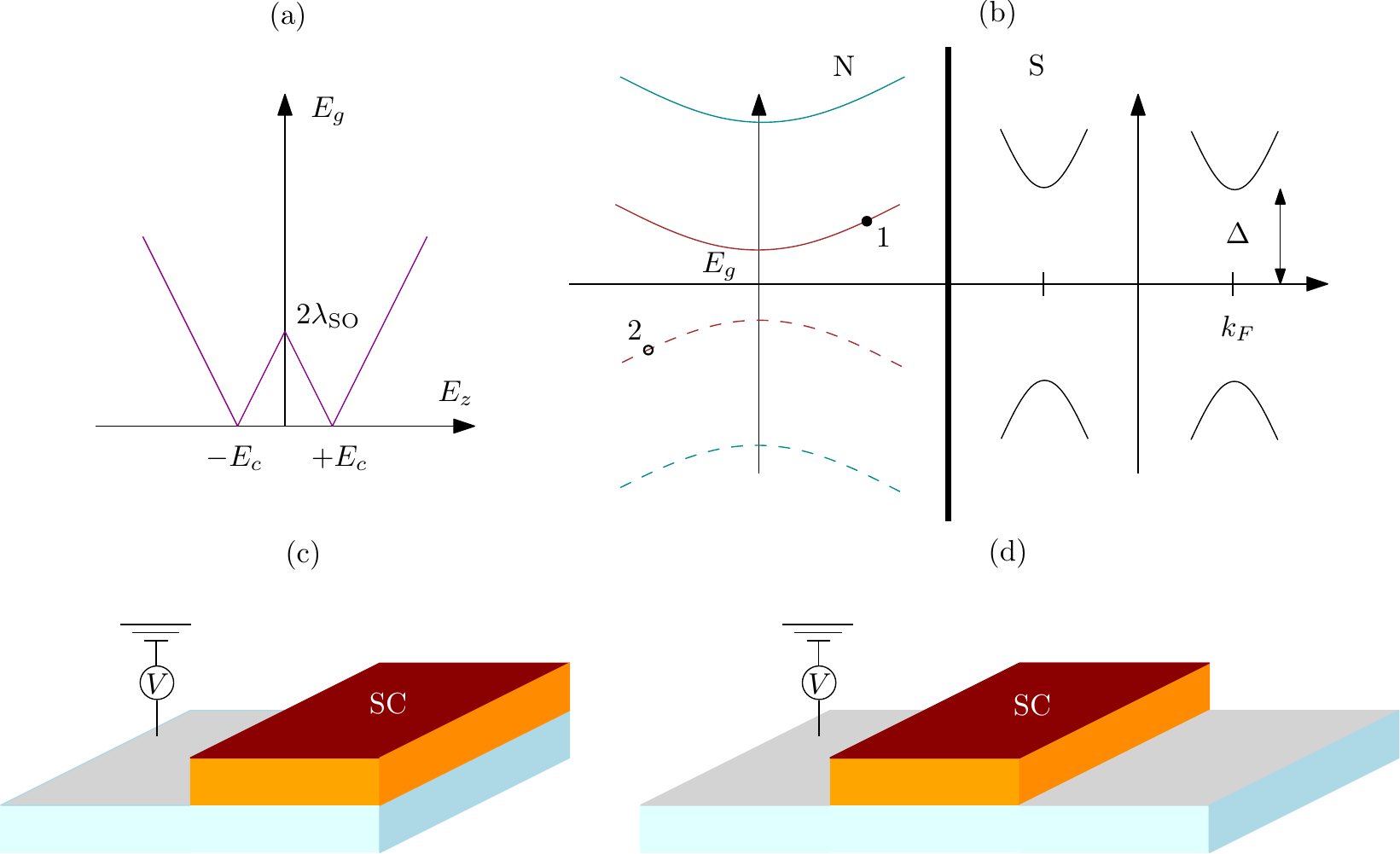}}
\caption{(Color online) (a) Plot of the insulating gap $E_g$ of normal-state silicene vs. an applied electric field $E_z$ perpendicular to the plane. (b) Band-structure in an N$\mid$S silicene junction. The two conduction bands are split at $\vk=0$. The process '1' indicates an incoming quasiparticle from one of the conduction bands, whereas '2' indicates the lost electron in the valence band when Andreev reflection occurs. (c) An effective  N$\mid$S silicene bilayer where superconductivity is induced via a proximate superconducting lead. (d) An effective N$\mid$S$\mid$N silicene junction to probe non-local transport. }
\label{fig:model} 
\end{figure}

One of the most recent advances in this field has been the synthesis of silicene \cite{silicene} which consists of silicon atoms arranged in a honeycomb pattern with a buckled sublattice structure. As in graphene, the states near the Fermi energy may be described by Dirac theory at two valleys $K$ and $K'$, but an important difference is that the fermions are massive in silicene due to a spin-orbit coupling which is much larger than in graphene. As a result, silicene is under the right circumstances a quantum spin Hall insulator with topologically protected edge states. In fact, it is possible \cite{ezawa} to achieve a rich variety of topological states in silicene due to a unique feature: the buckled structure causes the sublattices to respond differently to an applied electric field, which in turn induces a fermion mass-gap which is tunable. Closing and reopening this gap allows for a transition between different topological phases at a critical field value $|E_z|=E_c$ as shown in Fig. \ref{fig:model}(a).

The combination of a superconducting proximity effect with topologically protected edge-states is currently generating a lot of interest due to the possibility of creating Majorana fermions in this manner \cite{review_TI, majorana,Alicea,Beenakker,Leijnse}. However, there exists no study of proximity-induced superconductivity in silicene so far. In this Letter, we investigate precisely this topic and focus on the signature of Andreev reflection process, both locally and non-locally (which is usually dubbed crossed Andreev reflection (CAR)). We find that the possibility to tune both the band-gap via an electric field $E_z$ as well as the local Fermi level via a gate voltage provides an unparalleled control over the Andreev reflection process in silicene. In particular, we find that it is possible to generate a pure crossed Andreev reflection signal without any contamination from elastic co-tunneling. 
It is also shown that the supercurrent flowing in the superconductor$\mid$normal$\mid$superconductor (S$\mid$N$\mid$S) junction may be fully spin-valley polarized and that it is controllable by an external electric field.
This finding, combined with the observation that the Andreev reflection process is fully spin-valley polarized as will be described in detail later, demonstrates that silicene provides a unique environment for obtaining controllable superconducting transport with no counterpart in graphene or topological insulators. These results may pave the way for new perspectives for quantum transport polarized with novel degrees of freedom, namely the combined spin-valley product.

\textit{Theory and Results.} We consider a silicene layer made up by a buckled honeycomb lattice consisting of two sublattices $A$ and $B$ (see Fig. \ref{fig:model_app} in the Supplementary Information). Using a tight-binding
formalism, one obtains the following lattice Hamiltonian \cite{ezawa,Liu,Liu2}:
\begin{align}
H &= -t\sum_{\langle i,j \rangle,\alpha} c_{i\alpha}^\dag c_{j\alpha} + \frac{\i \lambda}{3\sqrt{3}} \sum_{\langle\langle i,j\rangle\rangle,\alpha,\beta} \nu_{ij} c_{\i\alpha}^\dag \sigma^z_{\alpha\beta} c_{j\beta} \notag\\
&+ l\sum_{i\alpha}\zeta_i E_z^i c_{i\alpha}^\dag c_{i\alpha} -\mu\sum_{i\sigma} c_{i\sigma}^\dag c_{i\sigma} + \sum_{i\sigma} (\sigma\Delta_0 c_{i\sigma}^\dag c_{i,-\sigma}^\dag + \text{h.c.})
\end{align}
Here, $t$ is the hopping element, $\lambda$ is the effective spin-orbit coupling parameter, $2l$ is the separation between the A- and B-sublattices in the $z$-direction, $E_z$ is an applied electric field, $\zeta_i=\pm1$ for the A (B) sites is the staggered sublattice potential term, while $\nu_{\ij} = (\mathbf{d}_i\times\mathbf{d}_j)/|\mathbf{d}_i\times\mathbf{d}_j|$ having defined $\mathbf{d}_i$ and $\mathbf{d}_j$ as the two nearest bonds connecting the next-nearest neighbors.
To describe quantum transport in the presence of a superconducting proximity effect, 
we derive an effective low-energy theory for excitations near the Dirac points $K_\eta$,
$\eta=\pm$. The details of this procedure are provided in the 
Supplementary Information. In the end, we obtain the following $\vk$-space Hamiltonian
when using a basis $\psi_\vk^\dag = [(\psi_{\vk,\sigma}^A)^\dag, (\psi_{\vk,\sigma}^B)^\dag, 
\psi_{-\vk,-\sigma}^A, \psi_{-\vk,-\sigma}^B]$:
\begin{align}
H_{\sigma,\mathbf{K}_\eta + \vk} &= \psi_\vk^\dag H_{\eta,\sigma}(\vk) \psi_\vk,\; H_{\eta,\sigma}(\vk) = \begin{pmatrix}
\hat{H}_0 & \sigma\Delta_0\hat{1} \\
\sigma\Delta_0^\dag\hat{1} & -\hat{H}_0\\
\end{pmatrix},\notag\\
\hat{H}_0 &= (lE_z-\eta\sigma\lambda_\text{SO})\hat{\tau}_z -\mu\hat{1} + v_F(\eta k_x \hat{\tau}_x - k_y\hat{\tau}_y)
\end{align}
with $\lambda_\text{SO}=\lambda/2$. Since we shall consider a hybrid junction consisting of normal silicene and a silicene-region
with proximity-induced superconductivity, it is instructive to discuss the eigenvalues and band-structure in these
regions separately. In the normal-state, silicene is an insulator with topological properties that may be controlled
by an external electric field as discussed previously. The excitation energies read: $E_{\eta,\sigma}(\vk) = \pm \sqrt{ k^2 + (l E_z - \eta\sigma\lambda_\text{SO})^2},$
having set the chemical potential $\mu_N=0$. The gap between conduction and valence band is then $E_g = 2|lE_z-\eta\sigma\lambda_\text{SO}|$ and we set $v_F=1$ for brevity of notation.

To allow for proximity-induced superconductivity in the region $x>0$, it is natural to include an electric doping level and thus a chemical potential $\mu_S \gg \lambda_\text{SO}, \Delta_0$ in order to have a
finite carrier-density at the Fermi level. The eigenvalues then read: $E_{\eta,\sigma}(\vk) = \pm \sqrt{ \Big(\sqrt{ k^2 + (lE_z - \eta\sigma\lambda_\text{SO})^2} \pm \mu_S \Big)^2 + |\Delta_0|^2}$.
It is now instructive to compare the band-structures in the N and S regions visually, as done in Fig. \ref{fig:model}(b). It is seen that in order for Andreev reflection to occur
the excitation gap in the N part must be smaller than the proximity-induced superconducting gap $\Delta_0$. In this way, an incoming electron-like quasiparticle from the N side
with energy $E$ (which must satisfy $E>E_g/2$ since there exists no states within the insulating gap) may be either normally reflected within the same conduction band or 
Andreev reflected. In the latter case, an electron of opposite spin is removed from the valence band and consequently the Andreev reflection process in undoped silicene is intrinsically
\textit{specular}: the Andreev-reflected hole has a group velocity parallel to its momentum.  

\begin{figure}[t!]
\centering
\resizebox{0.5\textwidth}{!}{
\includegraphics{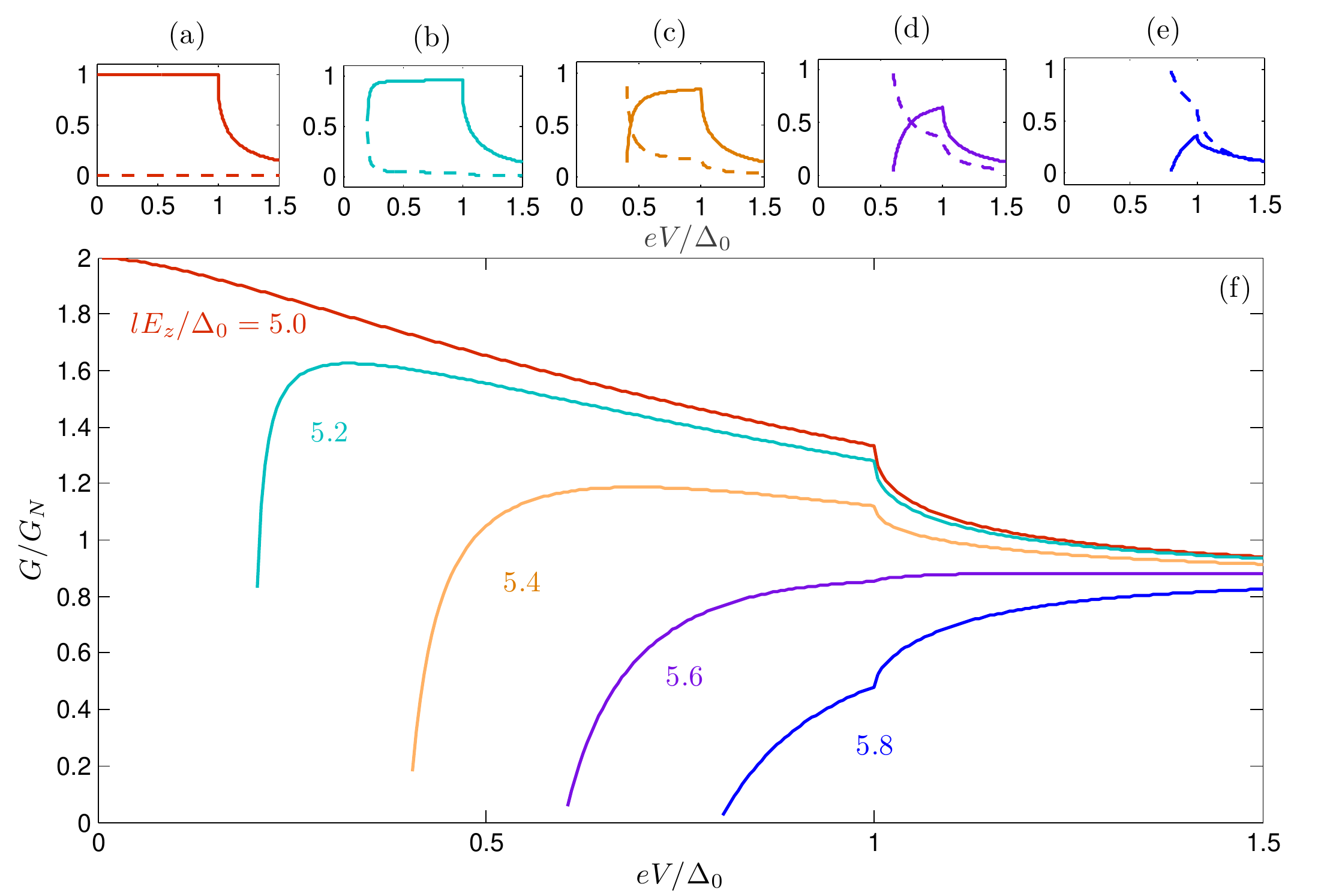}}
\caption{(Color online) (a)-(e): Normal (dashed lines) and Andreev reflection (full lines) probabilities for an N$\mid$S junction with $\eta=\sigma=+1$, $\lambda_\text{SO}/\Delta_0=5.0$, and $lE_z/\Delta_0$ ranging from 5.0 to 5.8 from (a) to (e). In (f), the conductance $G/G_N$ (averaged over spin and valleys) is plotted vs. bias voltage for the same choices of $lE_z/\Delta_0$.}
\label{fig:coeff} 
\end{figure}

Based on these observations, we are now in a position to write down the wavefunctions in the N and S regions as follows. 
At the interface $x=0$, we find that:
\begin{align}
\psi_N &= \frac{1}{\sqrt{2E\tau_+}}[\eta k_F\e{\i\eta\theta}, \tau_+,0,0] + \frac{r_e}{\sqrt{2E\tau_+}}[-\eta k_F\e{-\i\eta\theta},\tau_+,0,0]\notag\\
&+\frac{r_h}{\sqrt{2E\tau_-}}[0,0,\eta k_F\e{-\i\eta\theta},\tau_-],\notag\\
\psi_S &= \frac{t_e}{\sqrt{2}}[\eta\e{\i\eta\theta_S}u_+,u_+,\eta\e{\i\eta\theta_S}u_-\e{-\i\phi},u_-\e{-\i\phi}] \notag\\
&+\frac{t_h}{\sqrt{2}}[-\eta\e{-\i\eta\theta_S}u_-\e{\i\phi}, u_-\e{\i\phi}, -\eta\e{-\i\eta\theta_S}u_+,u_+].
\end{align}
The scattering coefficients $r_{\eta,\sigma}^e$, $r_{\eta,\sigma}^h$, $t_{\eta,\sigma}^e$, $t_{\eta,\sigma}^h$ denote normal reflection, Andreev reflection, and transmission as
electron- and hole-like quasiparticles, respectively.
The angle of incidence and transmission are related via $k_F\sin\theta = \mu\sin\theta_S$ where $k_F = \sqrt{E^2 - (lE_z-\eta\sigma\lambda_\text{SO})^2}$ and we have defined $\tau_\pm = E\pm(\eta\sigma\lambda_\text{SO} -lE_z)$ 
in addition to $u_\pm = [1/2 \pm \sqrt{E^2-\Delta_0^2}/2E]^{1/2}$. We note in passing that since the incident quasiparticles must have $E>(\eta\sigma\lambda_\text{SO} -lE_z)$ in order to exceed the insulating gap, $\tau_\pm$ is always real and positive. 
Since $\mu_S \gg k_F$, we may set $\theta_S=0$ for more transparent results. 

 The scattering coefficients may now be computed by matching the wavefunctions at the interface $x=0$ (as follows from conservation of current
flux, $\hat{v}_x\psi$ with $\hat{v}_x = \partial\hat{H}/\partial k_x$) and subsequently used
to find the conductance spectrum of the junction in the presence of an applied voltage: 
$G/G_N = \frac{1}{4} \sum_{\eta,\sigma} \int^{\pi/2}_{-\pi/2} \text{d}\theta \cos\theta (1 + |r_{\eta,\sigma}^h|^2 - |r_{\eta,\sigma}^e|^2)$. Note that an important difference from graphene is that we here cannot make
use of a valley degeneracy: the contribution to the charge conductance from each valley must be computed separately. 
From the boundary conditions, one then obtains an explicit analytical expression for the normal and Andreev reflection coefficients as follows:
\begin{align}\label{eq:coeff}
r_e = 2\cos\beta\Upsilon(\theta)\mathcal{D}^{-1},\; r_h = 4\e{\i(\eta\theta-\phi)}k_F\cos\theta\mathcal{D}^{-1},
\end{align}
with $\Upsilon(\theta)=\sum_\pm \pm\e{\pm\i\eta\theta}\tau_\mp $, $\mathcal{D} = 4(\i k_F\sin\beta\cos\theta + E\cos\beta)$ and $\e{\i\beta}=u_+/u_-$. As a consistency check, one may consider the "graphene" limit of the above results where $\lambda_\text{SO}=E_z=0$. In this case, we have $k_F=\tau_\pm = E$ so that one finds $r_h = \e{-\i(\beta+\phi)}$ for $\theta=0$. This agrees with the result of Ref. \cite{beenakker_prl_06} who found unity Andreev reflection probability even in the presence of a large Fermi vector mismatch (as in our case) for normal incidence. From the analytical expressions in Eq. (\ref{eq:coeff}), several observations can be made. Firstly, the Andreev reflection process is independent on whether $E_z<E_c$ or $E_z>E_c$ as long as the deviation $|E_z-E_c|$ from the critical field is the same. This may be seen by noting that these two regimes are related via the substitutions $\tau_+ \leftrightarrow \tau_-$ for which $|r_h|^2$ and $|r_e|^2$ are invariant. Secondly, it is seen that the probability for Andreev reflection, and thus the conductance of the junction, may be altered considerably by tuning the applied electric field $E_z$. We illustrate this in Fig. \ref{fig:coeff} setting $\lambda_\text{SO}/\Delta_0 = 5$. For the panels (a)-(e), we consider the Andreev (full lines) and normal (dashed lines) reflection probabilities as a function of bias voltage for normal incidence $\theta=0$. Due to the band splitting in the N part, only the $\eta=\sigma=+1$ and $\eta=\sigma=-1$ bands contribute to transport, and we consider in Fig. \ref{fig:coeff}(a)-(e) the $\eta=\sigma=+1$ case without loss of generality for an applied electric field $lE_z/\Delta_0$ ranging from $5.0$ to $5.8$. When the field is close to the critical one $E_c$, the Andreev reflection probability totally dominates normal reflection and one finds that it is unity for subgap energies exactly when $E_z=E_c$. Upon increasing the field $E_z$ and thus moving away from $E_c$, the normal reflection probability increases and eventually dominates Andreev reflection. Note that in each case, transport sets in only when $eV$ exceeds the insulating gap, the latter varying in magnitude with $E_z$. The experimental signature of this tunable Andreev reflection is seen in the conductance $G/G_N$ shown in (f): for fields close to $E_c$, the conductance is strongly enhanced at low bias voltages whereas it is suppressed at higher fields $E_z$ where normal reflection dominates. In effect, the applied electric field controls the Andreev reflection process and correspondingly the conductance of the junction, enabling a switching from Cooper pair transport to normal-state scattering.

\begin{figure}[t!]
\centering
\resizebox{0.5\textwidth}{!}{
\includegraphics{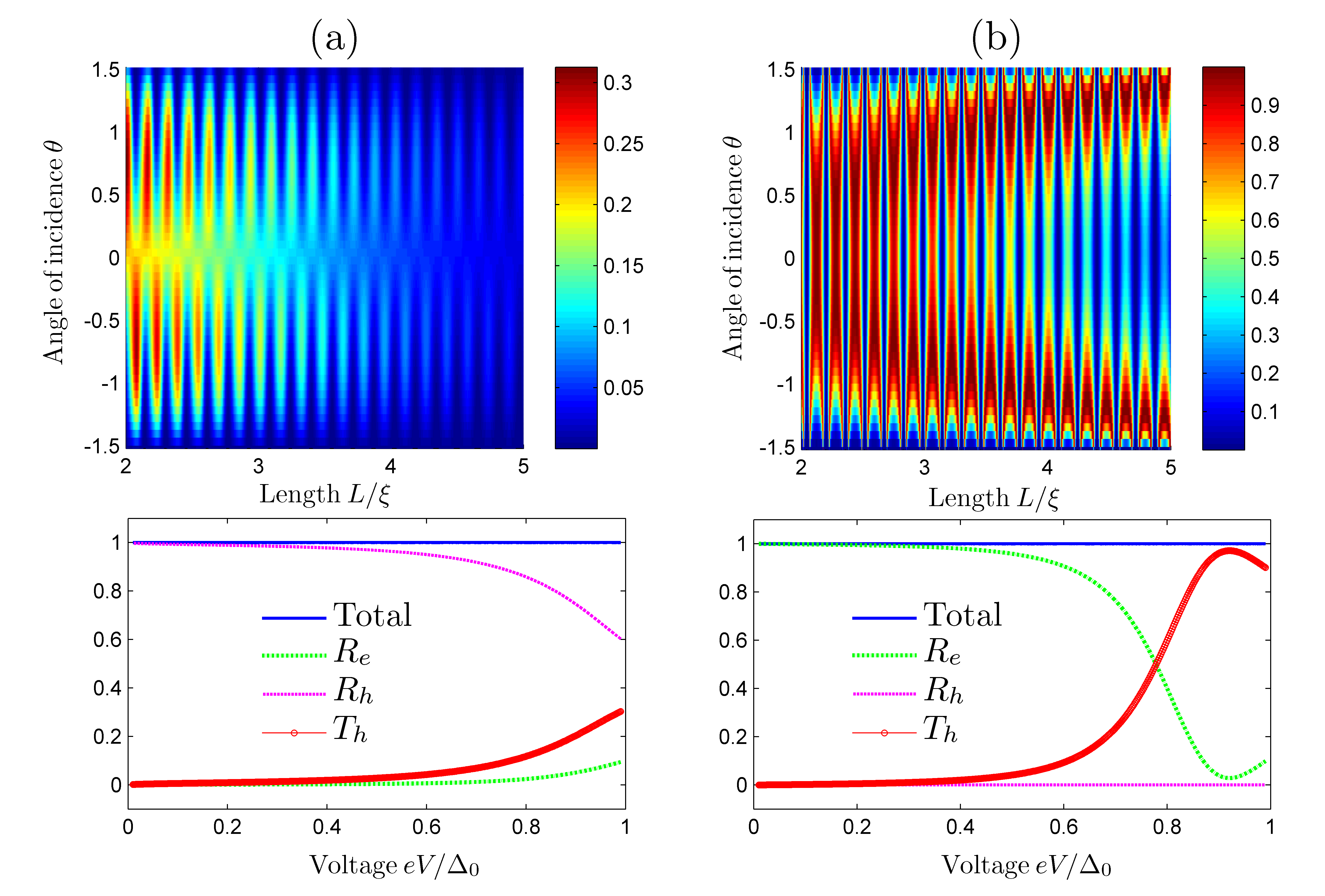}}
\caption{(Color online) \textit{Top row:} Contour-plot of the CAR probability for a bias voltage $eV/\Delta_0=0.9$ vs. angle of incidence $\theta$ and junction length $L$. \textit{Bottom row}: Probabilities for the different scattering events for a fixed junction length $L/\xi=2.1$ for normal incidence. In column (a), we consider scenario (i) as described in the main text (non-gapped/superconductor/gapped) whereas in (b) we consider scenario (ii) as described in the main text (gapped/superconductor/gapped). We have considered in all cases a strongly doped superconducting region with $\mu_S/\Delta_0=20$ and set $m_L/\Delta_0=0$ and $m_R/\Delta_0=5$ in (a) whereas $m_L/\Delta_0=m_R/\Delta_0=5$ in (b). The coefficients ($R_e, R_h, T_h$) are the probabilities for normal, Andreev, and crossed Andreev reflection, respectively. }
\label{fig:CAR} 
\end{figure}

We now demonstrate that silicene offers a unique testbed for probing non-local transport in the form of crossed Andreev reflection. The experimental setup is shown in Fig. \ref{fig:model}(d) and we assume as before a strongly doped superconducting region with large $\mu_S$. The fact that both the insulating gap and the Fermi level in silicene (due to its low density of states) may be controlled simply by external electric fields/gate voltages \cite{note_voltage}, is the key to obtaining not only a \textit{pure} CAR signal (without any elastic cotunneling) but also a non-local current which is \textit{fully spin-polarized} in each valley. To see how this may be obtained, let $2m_{L,R}$ denote the gap in the left and right normal silicene region between the lowest-lying conduction band and highest-lying valence band with $m=\lambda_\text{SO}-lE_z$. As in Fig. \ref{fig:model}(b), the two other bands are assumed to be separated largely and thus do not contribute to transport. Setting the Fermi level to the top of the valence band in the right region $(\mu_R = -m_R)$, the fate of non-local transport depends on the band-structure in the left region. We consider here two scenarios: (i) there is no gap in the left region $(m_L=0)$ with $\mu_L=0$, meaning that the electric field is equal to the critical value $E_c$, and (ii) there is a gap $2m_L$ in the left region and the Fermi level is tuned to lie right at the bottom of the conduction band $(\mu_L=m_L)$. In case (i), Andreev reflection can occur in addition to normal reflection for any incident energy since there is no gap in the spectrum whereas in the right region only CAR is possible. The reason is that an incident electron from the conduction band only has a gap to tunnel into in the right region. Consider now instead scenario (ii). In this case, local Andreev reflection is no longer possible since the spectrum is gapped on the left side. For the same reason as in case (i), elastic cotunneling is not possible either. This means that \textit{only} normal reflection and CAR are physically allowed scattering processes for this system. We emphasize here that it is not crucial that the Fermi level lies exactly on the gap edge, as considered above: a deviation from this simply means that the current-flow starts at a different applied voltage. We have chosen the above values to illustrate the principle used to obtain pure CAR as they offer the simplest visualization of the underlying idea.

The scattering probabilities are computed using the same method as in the N$\mid$S case, matching wavefunctions at the two interfaces with scattering coefficients $r_e$, $r_h$, $t_h$ associated with normal reflection, Andreev reflection, and crossed Andreev reflection. It is important to note that the belonging probability coefficients for each process ($R_e, R_h, T_h$) must be derived from the continuity equation, and are not necessarily equal to simply the modulus square of the above quantities; the interested reader may find the details of this calculation in the Supplementary Information. 
One obtains an expression for the (zero-temperature) non-local conductance $G_\text{nl}$ which may be experimentally measured:
\begin{align}\label{eq:nlcond}
\frac{G_\text{nl}}{G_0} & = \frac{1}{4}\sum_{\eta,\sigma}\int^{\pi/2}_{-\pi/2} \text{d}\theta \mathcal{P}_h|t_h|^2\sqrt{q_F^h-k_y^2},
\end{align}
where $\mathcal{P}_h= 1/(E-\mu_R)$, $q_F^h = \sqrt{(\mu_R-E)^2-m_R^2}$ is the wavevector of the CAR hole on the right side, $k_y = k_L\sin\theta$ is its transverse momentum, and $G_0$ is a normalization constant.
To investigate quantitatively the probabilities for these reflection processes to occur, consider Fig. \ref{fig:CAR}. We fix $\mu_S/\Delta_0=20$ and set the band gap to $m/\Delta_0=5$ when it is present in each region and also consider junction lengths $L\geq 2\xi$ where $\xi$ is the superconducting coherence length, since the non-selfconsistent approach used here is valid only for a sufficiently large superconducting region. In (a), the CAR process is shown both as a function of angle of incidence and junction length at a fixed voltage $eV/\Delta_0=0.9$ in the top panels and also as a function of bias voltage for a fixed junction length $L/\xi=2.1$ for normal incidence in the bottom panels. As seen, both local and non-local Andreev reflections are possible in this case and the maximum probability reached for the CAR process is about 30\% (we have verified this for other parameter choices). Still, it should be noted that CAR is the \textit{only} non-local transport process available due to the Fermi level lying right at the top of the valence band, which means that the current in the right N part is carried solely by crossed Andreev reflected holes. This is in complete contrast to usual metallic systems which typically gives the same order of magnitude for the probability of elastic cotunneling and CAR.

\begin{figure}[t!]
\centering
\resizebox{0.48\textwidth}{!}{
\includegraphics{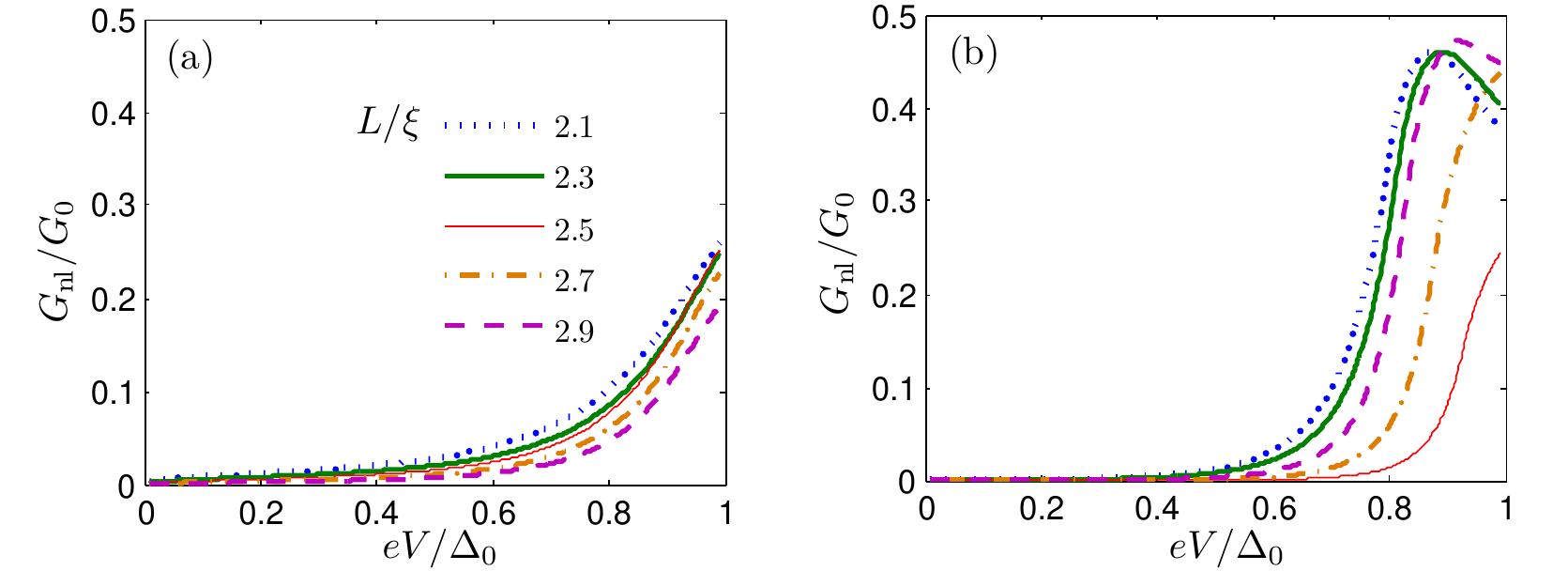}}
\caption{(Color online) Non-local conductance for (a) gapless and (b) gapped silicene on the left side [corresponding to scenarios (i) and (ii) described in the main text].}
\label{fig:CAR_cond} 
\end{figure}

The situation becomes even more intriguing when considering scenario (ii), where now CAR is the only physically allowed process in addition to normal reflection. In this case, CAR probability reaches essentially 100\% meaning that \textit{all} of the incoming electrons from the left N side combine with electrons from the right N side to produce Cooper pairs in the superconductor. A similar effect can be obtained at one specific voltage in graphene \cite{cayssol_prl_08}, but in that case elastic cotunneling occurs immediately upon deviating from that bias voltage. In the present case of silicene, there is no elastic cotunneling at all in the subgap regime and we have pure CAR at all voltages. In addition to generating a non-local Andreev (hole) current in this way, it is interesting to observe that this non-local current is fully spin-valley polarized. This means that in each valley, the current is fully spin-polarized with opposite spin-polarization in the two valleys such that the product spin$\otimes$valley is conserved. The non-local conductance defined in Eq. (\ref{eq:nlcond}) is shown in Fig. \ref{fig:CAR_cond} and is seen to show similar behavior to that of the CAR probability.

Finally, we have also computed the supercurrent flow through silicene, by considering an S$\mid$N$\mid$S junction. This setup is experimentally
viable and has previously been used to study the supercurrent through \eg graphene \cite{titov_prb_06, linder_prl_08} and topological insulators \cite{Fu,tanaka_prl_09, linder_prb_10}. We consider here ballistic transport which is appropriate
under the assumption of relatively short junction lengths satisfying $L\ll \xi$. In such a scenario,
the supercurrent is carried solely by Andreev bound-states (ABS) existing in the junction. These bound-states are formed by resonant Andreev-reflections occuring at the two interfaces and may be computed by setting up similar wavefunctions as in the N$\mid$S case and identifying the resonance energies. We compute the spin- and valley-dependent ABS energies $\varepsilon$ for a junction with a finite chemical potential $\mu_N$ in the normal region which is assumed to cross both of the conduction bands. Denoting the superconducting phase difference as $\Delta\phi$ and setting $\mu_N\gg\Delta_0$, we find that:
\begin{align}
\frac{\epsilon(\Delta\phi)}{\Delta_0} = \pm \sqrt{\frac{4\mathcal{M}^2\cos^2(\Delta\phi/2) + L^2(\mathcal{M}^2-k^2)^2}{4\mathcal{M}^2 + L^2(\mathcal{M}^2-k^2)^2}}
\end{align}
upon defining $\mathcal{M} = \mu_N + (\eta\sigma\lambda_\text{SO}-lE_z)$ and $k = \sqrt{\mu_N^2 - (\eta\sigma\lambda_\text{SO}-lE_z)^2}$. This gives rise to a supercurrent in the zero-temperature limit of the form:
\begin{align}
\frac{I(\Delta\phi)}{I_0} = \sum_{\eta\sigma} \frac{\mathcal{M}^2\sin\Delta\phi}{[4\mathcal{M}^2 + L^2(\mathcal{M}^2-k^2)^2]\epsilon(\Delta\phi)}.
\end{align}
The most interesting aspect of the above equation is that it explicitly depends on the applied electric field $E_z$, suggesting that one may experimentally control the supercurrent in a given sample by tuning the field $E_z$. Moreover, in the case were the Fermi level only crosses the lowest conduction band ($\eta=\sigma=\pm1$), the supercurrent is fully spin-valley polarized since Andreev reflection conserves this polarization. It would be interesting to compute the effect of a magnetic exchange field \cite{Yokoyama} on this result to see how it alters the spin-valley polarization of the supercurrent and if it is possible to obtain electrically controllable 0-$\pi$ oscillations, but we leave these issues for future investigations.

\textit{Acknowledgments.} 
This work was supported by Grant-in-Aid for Young Scientists (B) (No. 23740236) and the ``Topological Quantum Phenomena" (No. 25103709) Grant-in Aid for Scientific Research on Innovative Areas from the Ministry of Education, Culture, Sports, Science and Technology (MEXT) of Japan and the COST Action MP-1201 "Novel Functionalities through Optimized Confinement of Condensate and Fields".

\clearpage

\begin{widetext}
\begin{center}	
\textbf{SUPPLEMENTARY INFORMATION}
\end{center}
\begin{center}
\textit{\textbf{Derivation of low-energy Hamiltonian including superconductivity}}
\end{center}
The starting point is the lattice Hamiltonian in Ref. \cite{ezawa} where we add a superconducting pairing term:
\begin{align}
H = -t\sum_{\langle i,j \rangle,\alpha} c_{i\alpha}^\dag c_{j\alpha} + \frac{\i \lambda}{3\sqrt{3}} \sum_{\langle\langle i,j\rangle\rangle,\alpha,\beta} \nu_{ij} c_{\i\alpha}^\dag \sigma^z_{\alpha\beta} c_{j\beta} + l\sum_{i\alpha}\zeta_i E_z^i c_{i\alpha}^\dag c_{i\alpha} -\mu\sum_{i\sigma} c_{i\sigma}^\dag c_{i\sigma} + \sum_{i\sigma} (\sigma\Delta_0 c_{i\sigma}^\dag c_{i,-\sigma}^\dag + \text{h.c.})
\end{align}
We have added a chemical potential and superconducting pairing term compared to Ref.\cite{ezawa}. It is convenient to introduce separate fermion operators for the A and B sublattices. Moreover, we introduce the nearest-neighbor vectors $\delta^{A,B}_i$ and n.n.n vectors $\mathbf{a}_i$. Note that the n.n.n vectors are the same for both sublattices, but the n.n vectors are different. To be concrete, we have:
\begin{align}
\delta^A_1 = (a',0,a_z'),\; \delta^A_2 = (-a'/2, -a'\sqrt{3}/2, a_z'),\; \delta^A_3 = (-a'/2, a'\sqrt{3}/2, a_z'),\notag\\
\delta^B_1 = (a'/2,a'\sqrt{3}/2,-a_z'),\; \delta^B_2 = (a'/2, -a'\sqrt{3}/2, -a_z'),\; \delta^B_3 = (-a',0,-a_z').
\end{align}
and for the n.n.n vectors:
\begin{align}
\mathbf{a}_1 = (0,\sqrt{3}a',0),\; \mathbf{a}_2 = \frac{a'}{2}(3,\sqrt{3}),\; \mathbf{a}_3 = \frac{a'}{2}(3,-\sqrt{3}),\; \mathbf{a}_4 = (0,-\sqrt{3}a',0),\; \mathbf{a}_5 = \frac{a'}{2}(-3,-\sqrt{3}),\; \mathbf{a}_6 = \frac{a'}{2}(-3,\sqrt{3}).
\end{align}

\begin{figure}[hbt!]
\centering
\resizebox{0.23\textwidth}{!}{
\includegraphics{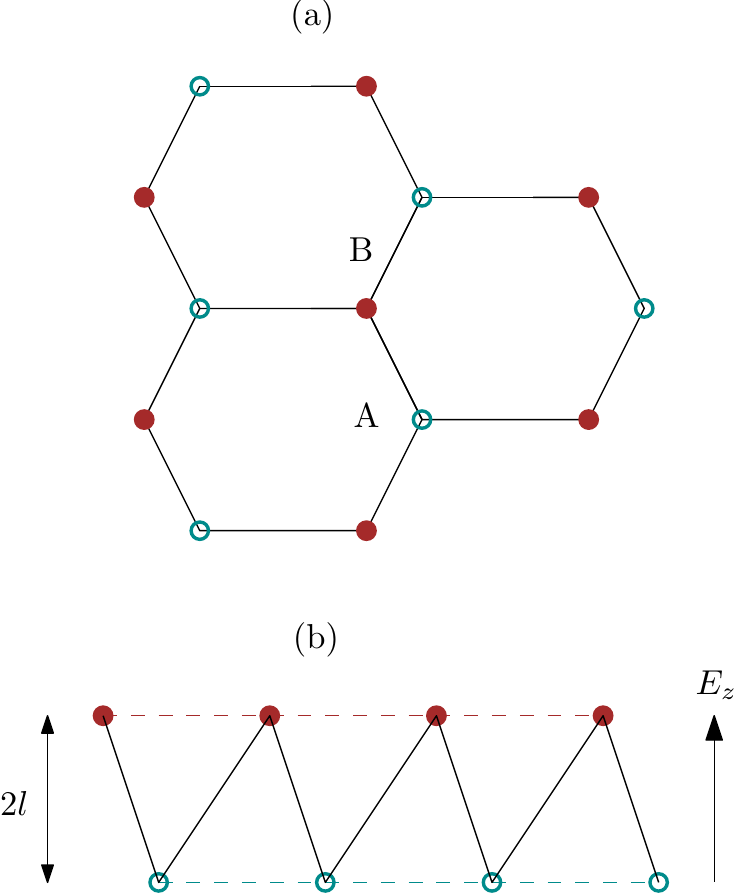}}
\caption{(Color online) (a) Top view of silicene consisting of two superpositioned triangular sublattices $A$ and $B$. (b) Side view of silicene in the presence of an external electric field $E_z$ perpendicular to the plane. }
\label{fig:model_app} 
\end{figure}

When inserting the sublattice operators and Fourier-transforming according to $A_{\vk\sigma} = \frac{1}{\sqrt{N_A}}\sum_i A_{i\sigma}\e{-\i\vk\mathbf{r}_i}$ (and similarly for B), the Hamiltonian reads:
\begin{align}
H &= -t\sum_{\vk,\delta^A,\sigma} A_{\vk\sigma} B_{\vk\sigma} \e{-\i\vk\delta^A} - t\sum_{\vk,\delta^B,\sigma}B_{\vk\sigma}^\dag A_{\vk\sigma} \e{-\i\vk\delta^B} + (lE_z-\mu)\sum_{\vk\sigma} A_{\vk\sigma}^\dag A_{\vk\sigma} - (lE_z+\mu)\sum_{\vk\sigma} B_{\vk\sigma}^\dag B_{\vk\sigma} \notag\\
&+\sum_{\vk\sigma}[\sigma\Delta_0 (A_{\vk\sigma}^\dag A_{-\vk,-\sigma}^\dag + B_{\vk\sigma}^\dag B_{-\vk,-\sigma}^\dag) + \text{h.c.}] \notag\\
&+\frac{\i\lambda}{3\sqrt{3}}\sum_{\alpha,\beta,\mathbf{a},\vk} (\nu^A_{\mathbf{a}}\e{-\i\vk\mathbf{a}} A_{\vk\alpha}^\dag \sigma_{\alpha\beta}^z A_{\vk\beta} + 
(\nu^B_{\mathbf{a}})\e{-\i\vk\mathbf{a}} B_{\vk\alpha}^\dag \sigma_{\alpha\beta}^z B_{\vk\beta})
\end{align}
where we have defined:
\begin{align}
\nu^A_{\mathbf{a}_2} = \nu^A_{\mathbf{a}_4} = \nu^A_{\mathbf{a}_6} = -\nu^A_{\mathbf{a}_1} = -\nu^A_{\mathbf{a}_1} = -\nu^A_{\mathbf{a}_5} = 1,\notag\\
\nu^B_{\mathbf{a}_2} = \nu^B_{\mathbf{a}_4} = \nu^B_{\mathbf{a}_6} = -\nu^B_{\mathbf{a}_1} = -\nu^B_{\mathbf{a}_1} = -\nu^B_{\mathbf{a}_5} = -1.
\end{align}
We now perform the summations over the n.n and n.n.n vectors and obtain:
\begin{align}
\gamma_\vk &\equiv \sum_{\delta^A} \e{-\i\vk\delta^A}= \Big(\sum_{\delta^B} \e{-\i\vk\delta^B}\Big)^*= \e{-\i k_xa'} + 2\cos(\sqrt{3}k_ya'/2)\e{\i k_xa'/2},\notag\\
2\i\eta_\vk &\equiv \sum_{\mathbf{a}} \nu^A_{\mathbf{a}} \e{-\i\vk\mathbf{a}} = - \sum_{\mathbf{a}} \nu^B_{\mathbf{a}} \e{-\i\vk\mathbf{a}} = 2\i[\sin(\sqrt{3}a'k_y) + \sin(3k_xa'/2 - \sqrt{3}k_ya'/2) - \sin(3k_xa'/2 + \sqrt{3}k_ya'/2)].
\end{align}
Since we are interested in the behavior near the Dirac points in the BZ, $\mathbf{K}_\pm = (0,\pm 4\pi/3\sqrt{3}a',0)$, we now expand the Hamiltonian for small $\vk$ around these points by letting $k_x \to k_x$ and $k_y \to \pm 4\pi/3\sqrt{3}a' + k_y$ where now $k_xa'\ll1$, $k_ya'\ll1$. Up to first order in $\vk$, we then find from the above definitions that:
\begin{align}
\eta_{\mathbf{K}_\pm + \vk} = \mp\frac{3\sqrt{3}}{2},\; \gamma_{\mathbf{K}_\pm + \vk} = -\frac{3a'}{2}(\i k_x\pm k_y).
\end{align}
It is useful to note the following identities: 
\begin{align}
\eta_{-\mathbf{K}_\pm - \vk} = \eta_{\mathbf{K}_\mp - \vk},\; \gamma_{-\mathbf{K}_\pm - \vk} = \gamma_{\mathbf{K}_\mp - \vk} 
\end{align}
which we shall use below. We can now rewrite Hamiltonian at the $\mathbf{q}\equiv \mathbf{K}_\pm + \vk$ point as follows
\begin{align}
H_{\mathbf{K}_\pm + \vk} &= \frac{3ta'}{2}\sum_\sigma\Big[ (\i k_x\pm k_y)A_{\vq\sigma}^\dag B_{\vq\sigma} + (-\i k_x\pm k_y)B_{\vq\sigma}^\dag A_{\vq\sigma} - (\i k_x\mp k_y)A_{-\vq\sigma}^\dag B_{-\vq\sigma} - (-\i k_x\mp k_y)B_{-\vq\sigma}^\dag A_{-\vq\sigma}\Big]\notag\\
&+ \frac{\epsilon_-}{2}\sum_\sigma(A_{\vq\sigma}^\dag A_{\vq\sigma} + A_{-\vq\sigma}^\dag A_{-\vq\sigma}) - \frac{\epsilon_+}{2}\sum_\sigma(B_{\vq\sigma}^\dag B_{\vq\sigma} + B_{-\vq\sigma}^\dag B_{-\vq\sigma})\notag\\
&+\frac{1}{2}\sum_\sigma[\sigma\Delta_0(A_{\vq\sigma}^\dag A_{-\vq,-\sigma}^\dag + B_{\vq\sigma}^\dag B_{-\vq,-\sigma}^\dag) + \sigma\Delta_0^\dag(A_{-\vq,-\sigma}A_{\vq\sigma} + B_{-\vq,-\sigma}B_{\vq\sigma})]\notag\\
&-\frac{\lambda}{3\sqrt{3}}\sum_{\alpha\beta}\Big[ (\mp3\sqrt{3}/2) (A_{\vq\alpha}^\dag \sigma_{\alpha\beta}^z A_{\vq\beta} - B_{\vq\alpha}^\dag \sigma_{\alpha\beta}^z B_{\vq\beta}) + (\pm3\sqrt{3}/2)(A_{-\vq\alpha}^\dag \sigma_{\alpha\beta}^z A_{-\vq\beta} - B_{-\vq\alpha}^\dag \sigma_{\alpha\beta}^z B_{-\vq\beta})\Big]
\end{align}
by making use of anticommutation relations and $\sum_\vk = \sum_{-\vk}$. We defined $\epsilon_\pm = lE_z\pm\mu$. Introducing the basis vector:
\begin{align}
\psi_\vq = [A_{\vq\uparrow}^\dag, B_{\vq\uparrow}^\dag, A_{-\vq\downarrow}, B_{-\vq\downarrow}, A_{\vq\downarrow}^\dag, B_{\vq\downarrow}^\dag, A_{-\vq\uparrow}, B_{-\vq\uparrow}]^\dag,
\end{align}
we can write the Hamiltonian in block-diagonal form as follows:
\begin{align}
H_{\mathbf{K}_\pm + \vk} = \psi_\vq^\dag M_\vq \psi_\vq,
\end{align}
where the matrix $M_\vq$ is:
\begin{align}
\begin{pmatrix}
\pm\lambda_\text{SO} + \epsilon_- & v_F(\i k_x\pm k_y) & \Delta_0 & 0 & 0 & 0 & 0 & 0\\
v_F(-\i k_x\pm k_y) & \mp\lambda_\text{SO}-\epsilon_+ & 0 & \Delta_0 & 0 & 0 & 0 & 0 \\
\Delta_0^\dag & 0 & \mp\lambda_\text{SO}-\epsilon_- & v_F(-\i k_x\mp k_y) & 0 & 0 & 0 & 0\\
0 & \Delta_0^\dag & v_F(\i k_x\mp k_y) & \pm \lambda_\text{SO} + \epsilon_+ & 0 & 0 & 0 & 0\\
0 & 0 & 0 & 0 & \mp \lambda_\text{SO}+\epsilon_- & v_F(\i k_x\pm k_y) & -\Delta_0 & 0 \\
0 & 0 & 0 & 0 & v_F(-\i k_x\pm k_y) & \pm \lambda_\text{SO} - \epsilon_+ & 0 & -\Delta_0\\
0 & 0 & 0 & 0 & -\Delta_0^\dag & 0 & \pm \lambda_\text{SO} - \epsilon_- & v_F(-\i k_x \mp k_y) \\
0 & 0 & 0 & 0 & 0 & -\Delta_0^\dag & v_F(\i k_x\mp k_y) & \mp\lambda_\text{SO}+\epsilon_+ \\
\end{pmatrix}
\end{align}
where we defined $\lambda_\text{SO} = \lambda/2$ and $v_F = 3ta'/4$. For easier comparison with Ref. \cite{ezawa}, consider the upper 4$\times$4 block and the $\mathbf{K}_+$ point with $\mu=0$:
\begin{align}
\begin{pmatrix}
\lambda_\text{SO} + lE_z & v_F(\i k_x\pm k_y) & \Delta_0 & 0 \\
v_F(-\i k_x\pm k_y) & -\lambda_\text{SO}-lE_z & 0 & \Delta_0 \\
\Delta_0^\dag & 0 & -\lambda_\text{SO}-lE_z & v_F(-\i k_x\mp k_y) \\
0 & \Delta_0^\dag & v_F(\i k_x\mp k_y) & +\lambda_\text{SO} + lE_z \\
\end{pmatrix}
\end{align}
One difference is that $k_x$ and $k_y$ are interchanged in comparison, but this can be changed by renaming our coordinate system $x\leftrightarrow y$ in which case we should also let $\lambda_\text{SO} \to -\lambda_\text{SO}$ since the determinant of the Jacobian for this transformation is -1 which alters the vector cross-products in the spin-orbit coupling term of the Hamiltonian. For compact notation, we can then write down a 4$\times$4 Hamiltonian valid for spin $\sigma$ and valley $\mathbf{K}_\eta$ where $\eta=\pm$. The basis is $\psi_{\vk\sigma}^\dag = [A_{\vk\sigma}^\dag, B_{\vk\sigma}^\dag, A_{-\vk,-\sigma}, B_{-\vk,-\sigma}]$ and the Hamiltonian is:
\begin{align}
H_{\vk\sigma} = \begin{pmatrix}
lE_z -\mu -\eta\sigma\lambda_\text{SO} & v_F(\eta k_x+\i k_y) & \sigma\Delta_0 & 0\\
v_F(\eta k_x-\i k_y) & -lE_z-\mu +\eta\sigma\lambda_\text{SO} & 0 & \sigma\Delta_0 \\
\sigma\Delta_0^\dag & 0 & -lE_z+\mu +\eta\sigma\lambda_\text{SO} & v_F(-\eta k_x-\i k_y) \\
0 & \sigma\Delta_0^\dag & v_F(-\eta k_x+\i k_y) & lE_z+\mu -\eta \sigma\lambda_\text{SO}
\end{pmatrix}
\end{align}
We obtain four possible eigenvalues for a given $\eta$ and $\sigma$, as usual:
\begin{align}
E_{\vk\sigma} = \pm \sqrt{\Big(\sqrt{v_F^2 k^2 + (lE_z-\eta\sigma\lambda_\text{SO})^2} \pm \mu\Big)^2 + |\Delta_0|^2}
\end{align}
The eigenvectors can now be identified by a direct product of the eigenvectors for the non-superconducting state and the usual BCS coherence factors. For instance, for the eigenvalue $E = \sqrt{\Big(\sqrt{v_F^2 k^2 + (lE_z+\lambda_\text{SO})^2} - \mu\Big)^2 + |\Delta_0|^2}$ we find that:
\begin{align}
\psi = \begin{pmatrix}
\frac{v_F(k_x+\i k_y)u}{\tau - \lambda_\text{SO} - \l E_z} \\
u\\
\frac{v_F(k_x+\i k_y)v\e{-\i\phi}}{\tau -\lambda_\text{SO}-l E_z}\\
v\e{-\i\phi}\\
\end{pmatrix}
\end{align}
where we defined for brevity's sake $\tau = \sqrt{v_F^2 k^2 + (lE_z+\lambda_\text{SO})^2}$ and
\begin{align}
u = \sqrt{\frac{1}{2}(1 + (\tau-\mu)/E)},\; v = \sqrt{\frac{1}{2}(1-(\tau-\mu)/E)}.
\end{align}
Above, $\phi$ is the superconducting phase. \\
\text{ }\\
\begin{center}
\textit{\textbf{Wavefunctions and probability coefficients for non-local transport}}
\end{center}
We now set $v_F=1$ for brevity of notation. Although it is in principle possible to write down a set of wavefunctions valid for an arbitrary parameter set, the corresponding analytical expressions are far too unwieldy to be of any use. Therefore, we concentrate on the scenarios (i) and (ii) in the N$\mid$S$\mid$N junctions treated in the main text. For those cases, the wavefunction in the left normal region reads:
\begin{align}
\psi_L = \frac{1}{\mathcal{N}_e}\begin{pmatrix}
\eta k_{x,e}^L +\i k_y\\
E + \mu_L + m_L\\
0\\
0\\
\end{pmatrix}\e{\i k_{x,e}^L x} + \frac{r_h}{\mathcal{N}_h} \begin{pmatrix}
0\\
0\\
\eta k_{x,h}^L - \i k_y\\
E-\mu_L-m_L\\
\end{pmatrix}\e{-\i k_{x,h}^L x} + \frac{r_e}{\mathcal{N}_e} \begin{pmatrix}
-\eta k_{x,e}^L +\i k_y\\
E + \mu_L + m_L\\
0\\
0\\
\end{pmatrix}\e{-\i k_{x,e}^L x}.
\end{align}
In the superconducting region one derives:
\begin{align}
\psi_S = t_1\begin{pmatrix}
\eta\e{\i\beta}\\
\e{\i\beta}\\
\eta\e{-\i\phi}\\
\e{-\i\phi}\\
\end{pmatrix}
\e{(\i\mu_S-\kappa)x} + t_2 \begin{pmatrix}
-\eta\e{\i\beta}\\
\e{\i\beta}\\
-\eta\e{-\i\phi}\\
\e{-\i\phi}\\
\end{pmatrix}\e{-(\i\mu_S-\kappa)x} +
t_3\begin{pmatrix}
-\eta\e{\i\phi}\\
\e{\i\phi}\\
-\eta\e{\i\beta}\\
\e{\i\beta}\\
\end{pmatrix}\e{-(\i\mu_S+\kappa)x} + 
t_4\begin{pmatrix}
\eta\e{\i\phi}\\
\e{\i\phi}\\
\eta\e{\i\beta}\\
\e{\i\beta}\\
\end{pmatrix}\e{(\i\mu_S+\kappa)x},
\end{align}
whereas in the right normal region we have:
\begin{align}
\psi_R = \frac{t_e}{\mathcal{M}_e} \begin{pmatrix}
\eta k_{x,e}^R + \i k_y\\
E+\mu_R+m_R\\
0\\
0\\
\end{pmatrix}\e{\i k_{x,e}^R x} 
+ \frac{t_h}{\mathcal{M}_h}\begin{pmatrix}
0\\
0\\
-\eta k_{x,h}^R -\i k_y\\
E -\mu_R-m_R\\
\end{pmatrix}\e{\i k_{x,h}^R x}
\end{align}
Here, $\kappa = \sqrt{\Delta_0^2-E^2}$. The quantities $\mathcal{N}_{e,h}$ and $\mathcal{M}_{e,h}$ are normalization constants:
\begin{align}
\mathcal{N}_e &= \sqrt{2(E+\mu_L)(E+\mu_L+m_L)},\; \mathcal{N}_h = \sqrt{|\eta k_{x,h}^L-\i k_y|^2 + (E-\mu_L-m_L)^2},\notag\\
\mathcal{M}_e &= \sqrt{|\eta k_{x,e}^R+\i k_y|^2 + (E+\mu_R+m_R)^2},\; \mathcal{M}_h = \sqrt{2(E-\mu_R)(E-\mu_R-m_R)}.
\end{align}
The wavevectors are given by 
\begin{align}
k_{x,(e,h)}^L &= \sqrt{k_F^{(e,h)}-k_y^2},\; k_F^{(e,h)} = \sqrt{(\mu_L\pm E)^2 -m_L^2},\notag\\
k_{x,(e,h)}^R &= \sqrt{q_F^{(e,h)}-k_y^2},\; q_F^{(e,h)} = \sqrt{(\mu_R\pm E)^2 -m_R^2},
\end{align}
with $k_y$ the transverse mode index. Matching the wavefunctions at each interface $x=0$ and $x=L$ and using the conservation of probability current, we find that the probability coefficients equal: 
\begin{align}
R_e &= |r_e|^2,\; T_e = 0,\; T_h = \frac{k_{x,h}^R}{k_{x,e}^L}\frac{(E+\mu_L)}{(E-\mu_R)}|t_h|^2, \notag\\
R_h &= \frac{\text{Re}\{k_{x,h}^L\}}{k_{x,e}^L}\frac{2(E+\mu_L)(E-\mu_L-m_L)}{|\eta k_{x,h}^L-\i k_y|^2 + (E-\mu_L-m_L)^2}|r_h|^2.
\end{align}
They satisfy the identity $R_e + R_h + T_e + T_h = 1$. 
\end{widetext}


\begin{thebibliography}{99}

\bibitem{review_graphene} A. H. Castro Neto \etal, Rev. Mod. Phys. \textbf{81}, 109 (2009).

\bibitem{review_TI} M. Z. Hasan and C. L. Kane, Rev. Mod. Phys. \textbf{82}, 3045 (2010).

\bibitem{nodal} It is appropriate to mention here the Dirac fermions emerging in the low-energy sector of the pseudogap phase for high-$T_c$ $d$-wave cuprates, which were discovered much earlier than either graphene or topological insulators.

\bibitem{silicene} B. Lalmi \etal, Appl. Phys. Lett. \textbf{97}, 223109 (2010); P.E. Padova \etal, Appl. Phys. Lett. \textbf{96}, 261905 (2010).

\bibitem{ezawa} M. Ezawa, New J. Phys. \textbf{14}, 033003 (2012); Phys. Rev. Lett. \textbf{109}, 055502 (2012).

\bibitem{majorana} X.-L. Qi and S.-C. Zhang, Rev. Mod. Phys. \textbf{83}, 1057 (2011).

\bibitem{Alicea} J. Alicea, Rep. Prog. Phys. \textbf{75}, 076501 (2012).

\bibitem{Beenakker} C. W. J. Beenakker,  Annu. Rev. Con. Mat. Phys. \textbf{4}, 113 (2013).

\bibitem{Leijnse} M. Leijnse and K. Flensberg, Semicond. Sci. Technol. \textbf{27}, 124003 (2012).

\bibitem{Liu} C.-C. Liu, W. Feng, and Y. Yao, Phys. Rev. Lett. \textbf{107}, 076802 (2011).

\bibitem{Liu2} C.-C. Liu, H. Jiang, and Y. Yao, Phys. Rev. B \textbf{84}, 195430 (2011).

\bibitem{beenakker_prl_06} C. Beenakker, Phys. Rev. Lett. \textbf{97}, 067007 (2006).

\bibitem{note_voltage} In general, $E_z$ and $\mu$ are both influenced when altering the external field, although the exact details should depend on the experimental setup such as the  substrate. 

\bibitem{cayssol_prl_08} J. Cayssol, Phys. Rev. Lett. \textbf{100}, 147001 (2008).

\bibitem{titov_prb_06}  M. Titov and C. W. J. Beenakker, Phys. Rev. B \textbf{74}, 041401 (2006).

\bibitem{linder_prl_08} J. Linder \etal, Phys. Rev. Lett. \textbf{100}, 187004 (2008).

\bibitem{Fu} L. Fu and C. L. Kane, Phys. Rev. B \textbf{79}, 161408(R) (2009).

\bibitem{tanaka_prl_09} Y. Tanaka \etal, Phys. Rev. Lett. \textbf{103}, 107002 (2009).

\bibitem{linder_prb_10} J. Linder \etal, Phys. Rev. B \textbf{81}, 184525 (2010).

\bibitem{Yokoyama} T. Yokoyama, Phys. Rev. B \textbf{87}, 241409(R) (2013).



\end{thebibliography}
\end{document}